# Structure and Electronic States of a Graphene Double Vacancy with an Embedded Si Dopant

Reed Nieman,[1] Adélia J. A. Aquino,[1,2] Trevor P. Hardcastle,[3,4] Jani Kotakoski,[5] Toma Susi,[5,a]

and Hans Lischka[1,2,a]

[1] Department of Chemistry and Biochemistry, Texas Tech University Lubbock, TX 79409-1061, USA

[2] School of Pharmaceutical Sciences and Technology, Tianjin University, Tianjin, 300072 P.R. China

[3] SuperSTEM Laboratory, STFC Daresbury Campus, Daresbury WA4 4AD, United Kingdom

[4] University of Leeds, School of Chemical and Process Engineering, Faculty of Engineering, 211 Clarendon Rd, Leeds LS2 9JT, United Kingdom

[5] Faculty of Physics, University of Vienna, Boltzmanngasse 5, A-1090 Vienna, Austria

# Abstract

Silicon represents a common intrinsic impurity in graphene, bonding to either three or four carbon neighbors respectively in a single or double carbon vacancy. We investigate the effect of the latter defect (Si-C$_4$) on the structural and electronic properties of graphene using density functional theory (DFT). Calculations based both on molecular models and with periodic boundary conditions have been performed. The two-carbon vacancy was constructed from pyrene (pyrene-2C) which was then expanded to circumpyrene-2C. The structural characterization of these cases revealed that the ground state is slightly non-planar, with the bonding carbons displaced from the plane by up to ±0.2 Å. This non-planar structure was confirmed by embedding the defect into a 10×8 supercell of graphene, resulting in 0.22 eV lower energy than the previously considered planar structure. Natural bond orbital (NBO) analysis showed sp$^3$ hybridization at the silicon atom for the non-planar structure and sp$^2$d hybridization for the planar structure. Atomically resolved electron energy loss spectroscopy (EELS) and corresponding spectrum simulations provide a mixed picture: a flat structure provides a slightly better overall spectrum match, but a small observed pre-peak is only present in the corrugated simulation. Considering the small energy barrier between the two equivalent corrugated conformations, both structures could plausibly exist as a superposition over the experimental timescale of seconds.

[a]Authors to whom correspondence should be addressed: hans.lischka@univie.ac.at, toma.susi@univie.ac.at

# I. INTRODUCTION

The modification of the physical properties of graphene sheets and nanoribbons, in particular the introduction of a band gap via chemical adsorption, carbon vacancies, and the incorporation of dopant species has gained a great deal of attention.[1, 2] Silicon is an interesting element as a dopant as it is isovalent to carbon. It is also a common intrinsic impurity in graphene sheets grown by chemical vapor deposition[3, 4] (CVD) and in graphene epitaxially grown by the thermal decomposition of silicon carbide[5, 6]. Further, dopant impurities have been shown to have significant effects on the transport properties of graphene.[7-9]

Silicon dopants in graphene exist predominantly in two previously identified forms: a non-planar, threefold-coordinated silicon atom in a single carbon vacancy, referred to as Si-$C_3$, and a fourfold-coordinated silicon atom in a double carbon vacancy, called Si-$C_4$, which is thought to be planar. An experimental work from 2012 by Zhou and coauthors[10] showed via simultaneous scanning transmission electron microscopy (STEM) annular dark field imaging (ADF) and electron energy loss spectroscopy (EELS) and, additionally, by density functional theory (DFT) calculations that the two forms of single silicon doped graphene differ energetically, and that the silicon atom in Si-$C_3$ has $sp^3$ orbital hybridization and that the silicon atom in Si-$C_4$ is $sp^2d$ hybridized. Similarly, EELS experiments combined with DFT spectrum simulations confirmed the puckering of the Si in the Si-$C_3$ structure and supported the planarity of the Si-$C_4$ structure.[11] However, a less satisfactory agreement of the simulated spectrum with the experimental one was noted in the latter case. The results suggested that d-band hybridization of the Si was responsible for electronic density disruption found at these sites. Experiments and dynamical simulations performed on these defects have demonstrated[12] how the Si-$C_4$ structure is formed when an adjacent carbon atom to the silicon is removed. However, it was concluded that the Si-$C_3$ structure is more stable and can be readily reconstructed by a diffusing carbon atom.

Past investigations of the electronic structure of graphene defects have shown the usefulness of the molecule pyrene as a compact representation of the essential features of single (SV) and double vacancy (DV) defects. Using this model, chemical bonding and the manifold of electronic states were investigated by multireference configuration interaction (MRCI) calculations.[13-16] In case of the SV defect,[13] the calculations showed four electronic states (two singlet and two triplet) within a narrow margin of 0.1 eV, whereas for the DV structure,[14] a gap of ~1 eV to the lowest excited state was found. For the DV defect, comparison with density functional theory using the hybrid Becke three-parameter Lee, Yang, and Parr (B3-LYP) density functional[17] showed good agreement with the MRCI results. Whereas for the covalently bonded Si-$C_3$ structure no further low-lying states are to be expected, the situation is different for the Si-$C_4$ defect since in this case an open-shell Si atom is inserted into a defect structure containing low-lying electronic states.

Based on these experiences, a pyrene model will be adopted also in this work to investigate the geometric arrangement and orbital hybridization around the silicon atom in a double carbon vacancy (FIG. 1). In a second step, a larger circumpyrene structure (FIG. 2) will be used to study the effect of increasing number of surrounding benzene rings. Two types of functionals will be employed, the afore-mentioned B3LYP and, for comparison, the long-range corrected Coulomb-attenuating B3LYP method (CAM-B3LYP).[18] Special emphasis will be devoted to verifying the electronic stability of the closed-shell wavefunction with respect to triplet instability[19, 20] and of the optimized structures with respect to structural relaxation. Finally, periodic DFT simulations of graphene with the Perdew-Burke-Ernzerhof (PBE) functional[21, 22] will be compared to the molecular calculations and an EELS spectrum simulated based on the computed Si-$C_4$ structure compared to an atomic resolution STEM/EELS measurement.

## II. COMPUTATIONAL DETAILS

Pyrene doped with a single silicon atom was investigated by first optimizing the structure of pristine pyrene using the density functional B3LYP[17] with the 6-31G*[23] basis set. The two interior carbon atoms were then removed and replaced with one silicon atom, referred to in this investigation as pyrene-2C+Si. This unrelaxed structure was subjected to a variety of closed-shell (restricted DFT, RDFT) and open-shell (unrestricted DFT, UDFT) geometry optimizations using the hybrid density functionals B3LYP and CAM-B3LYP in combination with the 6-311G(2d,1p)[24-28] basis set. To extend to a larger system, pyrene-2C+Si was surrounded by benzene rings (circumpyrene-2C) and this structure was optimized using the B3LYP/6-31G** approach. Finally, the geometry optimization of cirumpyrene-2C+Si was performed using the B3LYP functional and the 6-311G(2d,1d) basis.

To simulate the geometric restrictions present in a large graphene sheet, the Cartesian coordinates of all carbon atoms on the outer rims of pyrene-2C+Si (FIG. 1) and cirumpyrene-2C+Si (FIG. 2) that are bonded to hydrogen atoms were frozen during the course of geometry optimization to their values in the pristine structures. The coordinates of both pyrene-2C+Si and circumpyrene-2C+Si were oriented along the y-axis (the long axis) in the xy-plane.

The orbital occupations of the open-shell structures were determined using the natural orbital population analysis (NPA)[29] calculated from the total density matrix, while the bonding character at the silicon for each structure was investigated using natural bond orbital analysis.[30-34]

The discovered corrugated ground state structure of the Si-$C_4$ defect was then introduced into a periodic 10×8 supercell of graphene, and the structure relaxed to confirm that this corrugation is reproduced at that level of theory. Finally, the Si $L_{2,3}$ EELS response of the defect was simulated[11, 35] by evaluating the perturbation matrix elements of transitions from the Si 2p core states to the unoccupied states calculated up to 3042 bands, with no explicit core hole.[36] The resulting densities of state were broadened using the OptaDOS package[37] with a 0.4 eV Gaussian instrumental broadening and semi-empirical 0.015 eV Lorentzian lifetime broadening.

The Gaussian 09 (ref. [38]) and TURBOMOLE[39] program packages were used for the molecular calculations, and the GPAW[40] and CASTEP[41] packages for the graphene simulations.

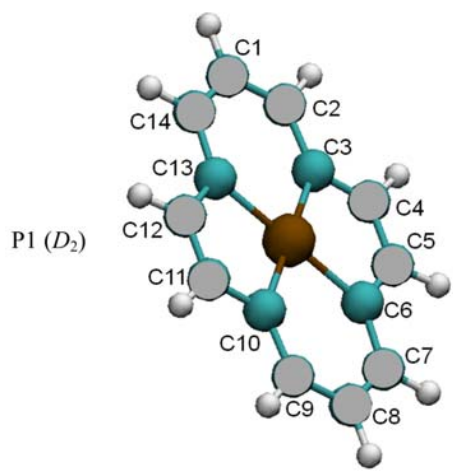

FIG. 1: Optimized pyrene-2C+Si structure P1 ($D_2$ symmetry) using the B3LYP/6-311G(2d,1p) method with frozen carbon atoms marked by gray circles

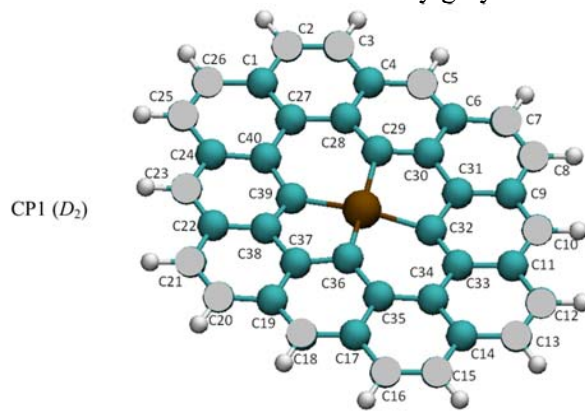

FIG. 2: Optimized circumpyrene-2C+Si structure CP1 ($D_2$ symmetry) using the B3LYP/6-311G(2d,1p) method with frozen carbon atoms marked by gray circles.

## III. RESULTS AND DISCUSSION

### A. The pyrene-2C+Si defect structure

The geometry of the unrelaxed, planar pyrene-2C+Si was first optimized using the B3LYP/6-311G(2d,1p) method at closed-shell level resulting in the $D_{2h}$ structure denoted P2. There were two imaginary frequencies present in the Hessian matrix. A displacement of the geometry of structure P2 along either of these imaginary frequencies led to a distortion of the planarity, lowering the point group to $D_2$ and producing the minimum energy structure P1, which was 0.736 eV lower in energy than structure P2 (TABLE I). The carbon atoms bonded to the silicon atom (atoms 3, 6, 10, and 13) were displaced from planarity by about ±0.2 Å while the silicon atom remained co-planar with the frozen carbon atoms on the periphery of pyrene-2C+Si. A similar tetrahedral geometry has been found for doping Al and Ga into DV graphene.[42] No triplet instability was found for either of the $D_{2h}$ and $D_2$ structures. Optimizing the planar pyrene-2C+Si structure at UDFT/B3LYP/6-311G(2d,1p) level for the high spin case resulted in structure P3, which was planar and belonging to the $D_{2h}$ point group. No imaginary frequencies were found in the Hessian matrix of structure P3; it is located 0.922 eV above the minimum energy structure P1.

TABLE I: Open-shell character, structural symmetry, spin multiplicity, bond distances (Å), and relative energies (eV) of selected optimized structures of pyrene-2C+Si.

| Structure | Closed/ Open Shell | Triplet instab. | #imag[a] | High/ Low Spin | Si[b]-C | C3-C13 | C3-C6 | ΔE |
|---|---|---|---|---|---|---|---|---|
| B3LYP/6-311G(2d,1p) | | | | | | | | |
| P1 ($D_2$)[c] | Closed | no | 0 | Low | 1.940 | 2.652 | 2.860 | 0.000 |
| P2 ($D_{2h}$) | Closed | no | 2 | Low | 1.932 | 2.634 | 2.826 | 0.736 |
| P3 ($D_{2h}$) | Open | - | 0 | High | 1.933 | 2.610 | 2.852 | 0.922 |

| | CAM-B3LYP/6-311G(2d,1p) | | | | | | | |
|---|---|---|---|---|---|---|---|---|
| P4 ($D_2$)[c] | Open | no | 0 | Low | 1.933[d] | 2.645[d] | 2.847[d] | 0.000 |
| P5 ($D_2$)[c] | Weakly open | no | 0 | Low | 1.936 | 2.648 | 2.852 | 0.040 |
| P6 ($D_2$)[c] | Closed | yes | 0 | Low | 1.938 | 2.648 | 2.857 | 0.056 |
| P7 ($D_{2h}$) | Open | no | 1 | Low | 1.923 | 2.616 | 2.820 | 0.602 |
| P8 ($D_{2h}$) | Closed | yes | 2 | Low | 1.931 | 2.631 | 2.828 | 0.734 |
| P9 ($D_2$)[e] | Open | - | 0 | High | 1.900 | 2.599 | 2.777 | 0.770 |
| P10 ($D_{2h}$) | Open | - | 1 | High | 1.900 | 2.598 | 2.772 | 0.826 |

[a]Number of imaginary frequencies. [b]Silicon always located in-plane. [c]The out-of-plane distance of carbon atoms that were not frozen in $D_2$ structures is ±0.2 Å [d]Averaged value of the four bonding carbons. [e]The out-of-plane distance of carbon atoms that were not frozen is ±0.07 Å

Structures P1-P3 were investigated also using the CAM-B3LYP/6-311G(2d,1p) method (TABLE I). For the planar geometry ($D_{2h}$ symmetry) two low spin structures were found. The first one was the closed-shell structure P8 that, however, turned out to be triplet instable. Re-optimization at UDFT/CAM-B3LYP level under $D_{2h}$ restrictions led to the open-shell structure P7 that was 0.13 eV more stable than the corresponding closed-shell structure. Structure P8 also had two imaginary vibrational frequencies, both of which were out-of-plane, leading to $D_2$ symmetry. Displacement along these modes and optimization at closed-shell level gave the non-planar structure P6 of $D_2$ symmetry. Similarly, following the out-of-plane imaginary frequency found in the Hessian of structure P7 led to a new structure P5 ($D_2$ symmetry) that was 0.016 eV lower in energy than structure P6. It had only a small open-shell character as can be seen from a natural orbital (NO) occupation of 0.089 and 1.911 $e$ respectively for the lowest unoccupied natural orbital (LUNO) and highest occupied natural orbital (HONO). This is much smaller than the NO occupation of the open-shell low spin structure P7 whose LUNO-HONO occupations were 0.302 and 1.698 $e$. Searching along the imaginary frequencies found for the previous structures and considering the triplet instability present in the wavefunctions led to structure P4 ($D_2$ symmetry), the new low spin, open-shell minimum-energy CAM-B3LYP structure which was 0.04 eV more stable than structure P5. The HONO-LUNO occupation for structure P4 was 1.899 *e* and 0.101 *e* respectively. Comparing the occupations from CAM-B3LYP to B3LYP shows that the range-corrected functional CAM-B3LYP presents a more varied picture than found with the B3LYP functional. In particular, open shell structures with non-negligible deviations of NO occupations from the closed shell reference values of two and zero, respectively, were found. However, the energy differences between these different states are quite small, only a few hundredths of an eV.

Structure P10 of TABLE I was obtained using the unrestricted high spin approach in the geometry optimization. The Hessian contained one imaginary frequency, an out-of-plane bending mode leading to $D_2$ symmetry. Following this imaginary frequency led to structure P9 that was about 0.06 eV more stable than structure P10. The bonding carbons of this new high spin structure were slightly out-of-plane by 0.07 Å.

Selected bond distances of the silicon-2C+Si complex are given in TABLE I for the B3-LYP and CAM-B3LYP results. Looking first at the distances from the silicon atom to its bonding carbon atoms computed at B3LYP/6-311G(2d,1p) level, the Si-C bond distances are found to vary by less than 0.01 Å in the different structures. The $D_2$ structure has the longest Si-C bond distance as would be expected from moving the bonding carbon atoms out-of-plane from the $D_{2h}$ to $D_2$ symmetry. The internuclear distance between adjacent, non-bonded carbon atoms C3 and C13 along the direction perpendicular to the long axis is 2.65 Å for structure P1, which is the largest among the three structures. This is again a result of the out-of-plane character of structure P1. The distance C3-C6 along the long axis of pyrene is ~0.2 Å longer than the perpendicular distance, reflecting the restrictions imposed by freezing the carbon atoms which would be connected to the surrounding graphene sheet (FIG. 1).

Next, we characterize the structures computed with the CAM-B3LYP/6-311G(2d,1p) method. The largest C3-C13 bond distance (TABLE I) is 2.65 Å for the low spin structures P5 and P6 ($D_2$), while the smallest, 2.60 Å, is found for the high spin, open-shell structure P10.

The dependence of the Si-C distances on the different structures and spin states is not very pronounced, though the high spin planar structures, structures P9 and P10, were found to have the smallest Si-C and C3-C13 distances; the same situation that was seen using B3LYP/6-311G(2d,1p).

Comparing the two above methods for pyrene-2C+Si, the most noticeable effect is that there is greater open-shell character for the low spin structures when using the CAM-B3LYP functional. The Si-C bond distance and C3-C13 intranuclear distances are also slightly smaller when using the CAM-B3LYP functional. But overall, agreement between the two methods is quite good.

## B. The circumpyrene-2C+Si defect structure

Pyrene-2C+Si was then surrounded by benzene rings to create the larger circumpyrene-2C+Si structure (FIG. 2) as a better model for embedding the defect into a graphene sheet. The edge carbon atoms linked to hydrogen atoms were again frozen. Description of the geometry of circumpyrene-2C+Si is presented in TABLE II for the B3LYP/6-311G(2d,1p) method. The geometry of the planar, unrelaxed circumpyrene-2C+Si was optimized using a closed-shell approach that led to the $D_{2h}$ structure CP2, which had one out-of-plane imaginary frequency. Following this mode, the non-planar $D_2$ symmetric structure CP1 was obtained that contained no imaginary frequencies. Optimizing the geometry of the planar, unrelaxed circumpyrene-2C+Si using an open-shell, high spin approach resulted in a geometrically stable (no imaginary frequencies) planar structure CP3 with $D_{2h}$ symmetry. The wave functions of circumpyrene-2C+Si structures CP1 and CP2 were triplet stable.

TABLE II. Structural symmetry, open-shell character, spin multiplicity, bond distances (Å), and relative energies (eV) of selected optimized structures of circumyrene-2C+Si.

| Structure | Closed/Open Shell | Triplet instab. | #imag[a] | High/Low Spin | Si[b]-C | C29-C39 | C29-C32 | ΔE |
|---|---|---|---|---|---|---|---|---|
| B3LYP/6-311G(2d,1p) | | | | | | | | |
| CP1 ($D_2$)[c] | Closed | no | 0 | Low | 1.909 | 2.690 | 2.741 | 0.000 |
| CP2 ($D_{2h}$) | Closed | no | 1 | Low | 1.902 | 2.668 | 2.713 | 0.118 |
| CP3 ($D_{2h}$) | Open | - | 0 | High | 1.912 | 2.662 | 2.745 | 1.254 |

[a]Number of imaginary frequencies. [b]Silicon always stays in-plane [c]The out-of-plane distance of non-frozen carbon atoms is ±0.2 Å in structures that are $D_2$ symmetric

The Si-C bond distances for the three investigated structures are very similar (TABLE II). The C29-C39 distance is largest for the non-planar, closed-shell structure P1 and smallest for the planar structures P2 and P3, but the differences are less than 0.03 Å.

The difference in energy between the non-planar and planar low spin structures decreases markedly when the 2C+Si defect is embedded in a larger hexagonal sheet (compare ΔE values in TABLE I and TABLE II). The energy difference decreases by ~0.6 eV from 0.74 eV in pyrene-2C+Si to 0.12 eV in circumpyrene-2C+Si. Also noteworthy, when comparing pyrene-2C+Si and circumpyrene-2C+Si is that the Si-C bond distances in circumpyrene-2C+Si are shorter by about 0.03 Å on average for the low spin cases. The differences in the internuclear distance between adjacent, non-bonded carbon atoms C29 and C39 (perpendicular to the long axis) and C29-C32 (parallel) is much less pronounced than in the pyrene case, which is a consequence of the more flexible embedding into the carbon network where none of the four C atoms surrounding Si is bonded to another atom which is frozen.

## C. The graphene-2C+Si defect structure

To confirm the obtained geometry and to validate that periodic DFT simulations reproduce the observed ground state structure, the Si-C$_4$ defect with an alternating ±0.2 Å corrugation of the four C atoms was placed in a 10×8 supercell of graphene and the structure and cell relaxed using the PBE functional. The structural optimization preserved the out-of-plane corrugation, and resulted in a total energy 0.22 eV lower than the flat structure. Thus, standard DFT is able to find the correct, non-planar structure when initializing away from the flat geometry. Further, a nudged elastic band calculation[43] shows that the two equivalent conformations (alternating which C atoms are up and which are down) are separated only by this energy barrier, leading to a Boltzmann factor of $1.7 \times 10^{-4}$ at room temperature and thus a rapid oscillation between the two equivalent structures for any reasonable vibration frequency.

Bader analysis[44] of the charge density shows that the Si-C bonds are rather polarized, with the Si donating 2.61 electrons shared by the four neighboring carbons. This is even larger than the bond polarization in 2D-SiC, where a Si donates 1.2 electrons (although each C has three Si neighbors, bringing the total to 0.4 per C instead of 0.65 here).[45]

### D. Natural bond orbital analysis

Natural bond orbital (NBO) analysis was used to further characterize the bonding of pyrene-2C+Si and circumpyrene-2C+Si and to describe in detail the charge transfer from silicon to carbon in the various structures.

The NBO analysis of pyrene-2C+Si is presented in TABLE III. The linear combination factor of the Si is less than half of that of the bonding C, a sign of the greater electronegativity of carbon compared to silicon. The non-planar, low spin, closed-shell structure P1 ($D_2$ symmetry) shows $sp^3$ orbital hybridization at the silicon and $sp^3$ orbital hybridization at the bonding carbon, consistent with the slight tetrahedral arrangement about the silicon center. The planar closed-shell structure P2 shows $sp^2d$ hybridization at the silicon center, consistent with previous experimental and computational interpretations[10], and approximately $sp^3$ hybridization at the bonding carbons. The high spin structure P3, which also has $D_{2h}$ symmetry, shows also sp²d hybridization at the silicon. Compared to B3LYP (structures P1-P3), bonding from CAM-B3LYP (TABLE SI of the Supplementary Information) is essentially the same.

TABLE III: NBO bonding character analysis of pyrene-2C+Si using B3LYP/6-311G(2d,1p). The discussed carbon atom is one of the four bonded to silicon.

| Structure | Bonding Character |
|---|---|
| | B3LYP/6-311G(2d,1p) |
| P1 ($D_2$) closed sh. | $0.54(sp^{2.98}d^{0.02})_{Si}+0.84(sp^{3.24}d^{0.01})_C$ |
| P2 ($D_{2h}$) closed sh. | $0.46(sp^{2.00}d^{1.00})_{Si}+0.89(sp^{2.97}d^{0.01})_C$ |
| P3 ($D_{2h}$) open sh., high spin | α: $0.46(sp^{2.00}d^{1.00})_{Si}+0.89(sp^{2.81}d^{0.01})_C$<br>β: $0.55(sp^{2.00}d^{1.00})_{Si}+0.83(sp^{3.61}d^{0.01})_C$ |

The NBO analysis of circumpyrene-2C+Si is presented in TABLE IV. The bonding character of structure CP1 is essentially sp³ for both the silicon and the bonding carbon as in structure P1 of pyrene-2C+Si, although the p orbital character on the carbon atom of this structure is reduced by about 0.60 $e$. Structure CP2 presents an interesting case, as no bonding NBOs were found. Instead, a series of four unoccupied valence lone pairs (LP*), one of s orbital character and three of p character, were located on the silicon and a corresponding set of occupied valence lone pairs (LP), one for each bonding carbon, was found with $sp^{3.58}d^{0.01}$ character. It is noted that the Si-C bonding character of the high spin structure CP3 changes markedly between the alpha shell and beta shell. In the alpha shell, the silicon and carbon atoms both exhibit sp¹ orbital hybridization with no d orbital contribution on the silicon, but in the beta shell, the silicon atom shows sp²d hybridization and the carbon atoms show sp³ hybridization as in the other planar cases.

The shapes of the NBOs describing the bond between Si and C is depicted in FIG. 3 for structures P1 ($D_2$ symmetry) and P2 ($D_{2h}$ symmetry) using the B3LYP/6-311G(2d,1p) method. They appear very similar in spite of the different hybridization on Si (sp³ vs. sp²d, TABLE III).

The plots of the NBOs for all other structures, even those for circumpyrene-2C+Si, are almost identical. The exception is structure CP2 of circumpyrene-2C+Si in which no Si-C bonds were found during the NBO analysis. The NBOs shown in FIG. S1 for this structure consist of the valence lone pairs found on each carbon atom (TABLE IV) possessing $sp^{3.58}$ orbital hybridization with a lobe of electron density found facing the silicon center. The NBOs of silicon are low-occupancy valence lone pairs, representing one s and three p NBOs. Even though this classification of bonding by means of lone pairs according to the NBO analysis looks quite different, the combination of these lone pairs is not expected to look substantially different from the localized bonds shown in the other structures, especially considering the strong polarity of the silicon-carbon bonds as discussed below.

TABLE IV: NBO bonding character analysis of circumpyrene-2C+Si. The discussed carbon atom is the one of the four bonded to silicon.

| Structure | Bonding Character |
|---|---|
| | B3LYP/6-311G(2d,1p) |
| CP1 ($D_2$) closed sh. | $0.51(sp^{2.96}d^{0.04})_{Si}+0.86(sp^{2.62}d^{0.01})_{C}$ |
| CP2 ($D_{2h}$) closed sh. | Four LP*: one $[1.0(s)_{Si}]$ + three $[1.00(p)_{Si}]$[a]  <br> four$[1.00(sp^{3.58}d^{0.01})_C]$[b] |
| CP3 ($D_{2h}$) open sh., high spin | α: $0.68(sp^{1.00}d^{0.00})_{Si}+0.74(sp^{1.00}d^{0.00})_C$  <br> β: $0.55(sp^{2.00}d^{1.00})_{Si}+0.84(sp^{3.41}d^{0.01})_C$ |

[a]Four unfilled valence lone pairs (LP*) on silicon (occupation less than 1 $e$)
[b]One occupied valence lone pair on each bonding carbon

Natural charges are instructive for illustrating the polarity of the silicon-carbon bonds and the charge transfer between silicon and the surrounding graphitic lattice. These are shown in FIG. 4 for circumpyrene-2C+Si (Structures CP2 and CP3) using the B3LYP/6-311G(2d,1p) method. The silicon center is strongly positive in all cases and more positive in circumpyrene-2C+Si for the low spin cases (1.7–1.8 $e$, structures CP1 and CP2) than the high spin case (1.4 $e$, structure

CP3). It thus seems that the Bader analysis presented above overestimates the charge transfer by over 40 %, despite being qualitatively correct. The bonding carbon in each case shows a larger negative charge compared to the remaining carbon atoms of circumpyrene. The charge on the bonding carbon in the low spin structures CP1 and CP2 is more negative than that of the high spin structure CP3.

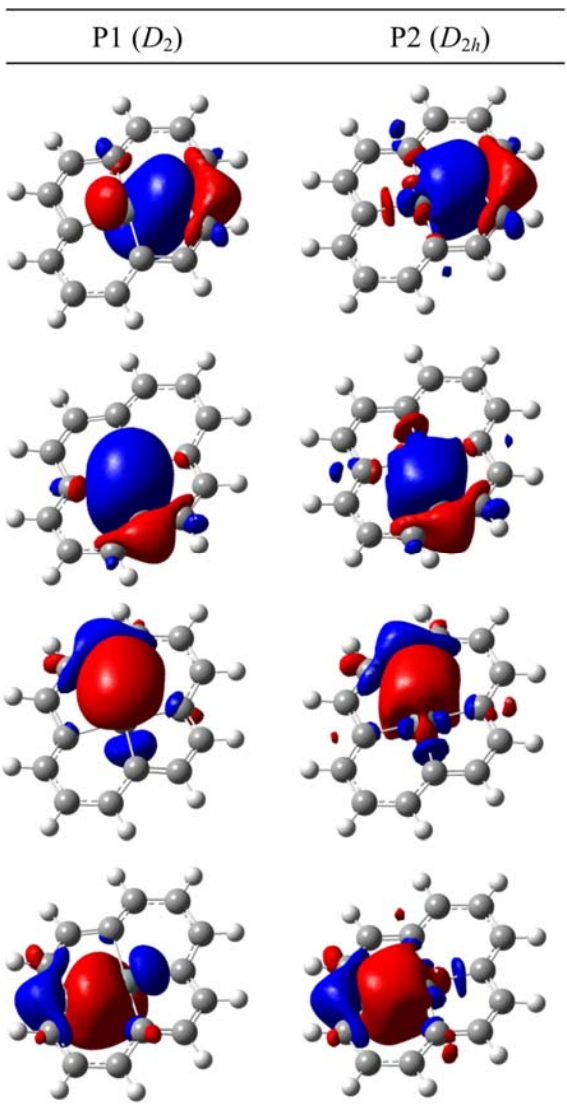

FIG. 3: NBO plots for pyrene-2C+Si using the B3LYP/6-311G(2d,1p) approach for structures P1 and P2. Isovalue = ±0.02 $e$/Bohr$^3$

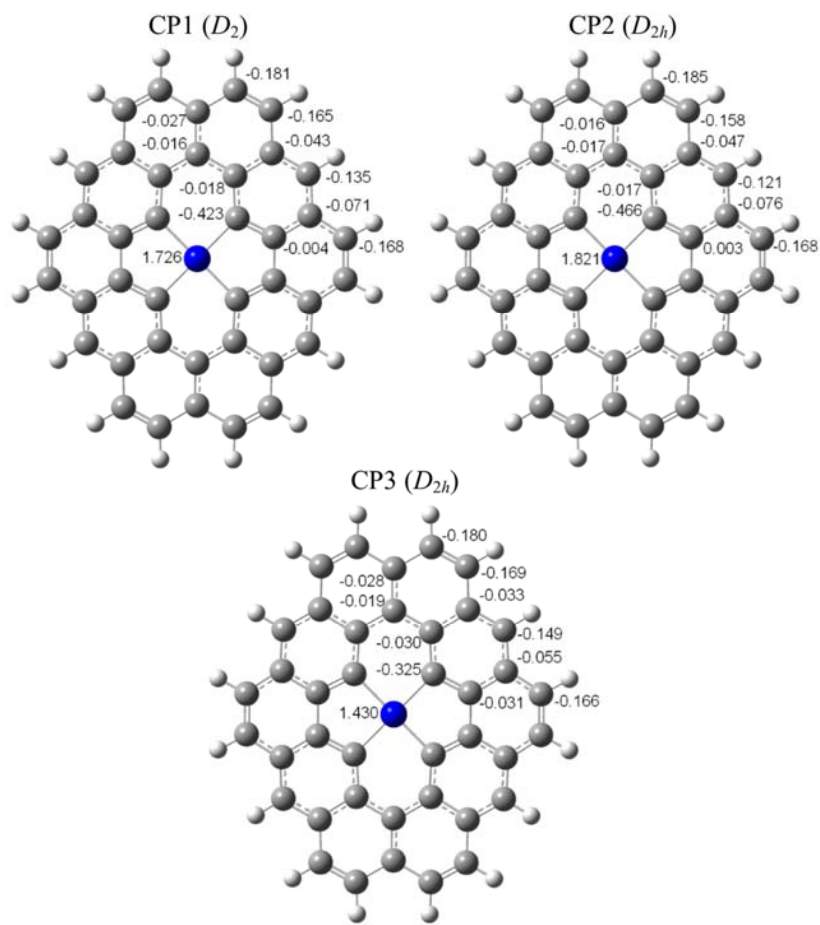

FIG. 4: Natural charges (*e*) of carbon and silicon in circumpyrene-2C+Si using the B3LYP/6-311G(2d,1p) method.

The natural charges of the remaining carbon atoms in circumpyrene-2C vary depending on whether they are bonded to other carbon atoms or to hydrogen. For the low spin structures CP1 and CP2, the natural charges on C atoms on the periphery of circumpyrene bonded to other C atoms is about one-third as negative as the natural charges of C atoms bonded to H atoms, while the natural charges of carbon atoms in the interior of circumpyrene and not bonded to Si are much smaller. Taken together, the magnitude of the Si-graphene charge transfer can be seen to be largest for the planar, closed-shell, low spin structure CP2, and smallest for the planar, high spin structure CP3.

The natural charges for pyrene-2C+Si using B3LYP/6-311G(2d,1p) are plotted in FIG. S2. The low spin structures P1 and P2 have respective positive charges on the silicon atom of 1.596 and 1.953 $e$, a larger difference than in the low spin structures CP1 and CP2 of circumpyrene-2C+Si where the silicon atom has positive charges of 1.726 and 1.821 $e$. Correspondingly, the difference in the negative charges of the bonding carbon atoms is also larger between these two structures than in circumpyrene. For the other carbons in pyrene, the magnitude of their charges varies less than in circumpyrene. For pyrene-2C+Si using the CAM-B3LYP/6-311G(2d,1p) method, the natural charges are plotted in FIG. S3. The same trend is seen for this method as in our other cases, namely that the closed-shell, low spin, planar structure P8 has the largest positive charge on silicon and the largest negative charge on the bonding carbon compared to the other structures.

### E. Molecular Electrostatic Potential

The molecular electrostatic potential (MEP) was computed for each structure and is presented in FIG. **5** for circumpyrene-2C+Si using the B3LYP/6-311G(2d,1p) method. The plots illustrate the natural charge transfer discussed previously: for structures CP1-CP3, the positive potential at the silicon corresponds to the positive charge buildup as discussed above. This positive potential at the silicon is more positive for structures CP1 and CP2 than for CP3 following the trend of the natural charges. There is also a negative potential diffusely distributed on the carbon atoms in circumpyrene.

The MEP plots for pyrene-2C+Si using the B3LYP/6-311G(2d,1p) method is shown in FIG. S4. Similar results are obtained when using the CAM-B3LYP functional (shown in FIG. S5). Finally, the low spin planar structures P2, P7, and P8 have a much more positive potential than either the non-planar structures or the high spin planar structures, represented by the deeper blue coloring in the figures.

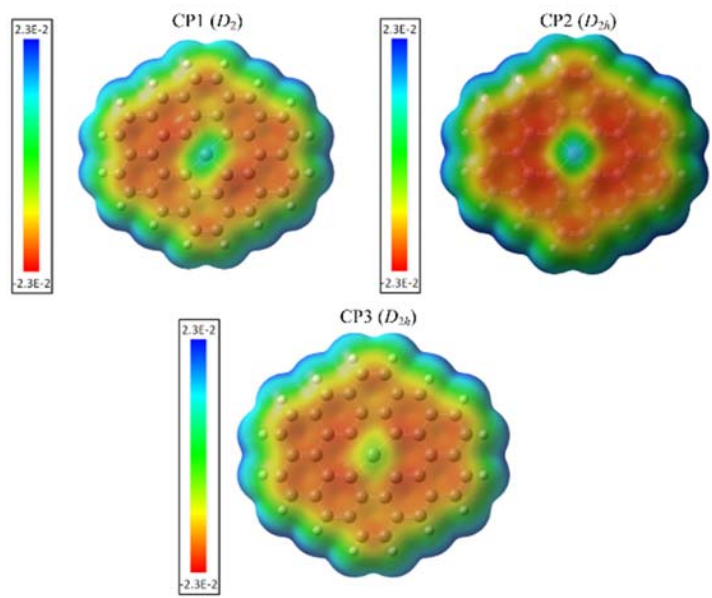

FIG. 5: Molecular electrostatic potential for circumpyrene-2C+Si mapped onto the electron density isosurface with an isovalue of 0.0004 $e$/Bohr$^3$ (B3LYP/6-311G(2d,1p)).

## F. Electron energy loss spectrum

The Si-C$_4$ defect in graphene was originally identified through a combination of atomic resolution STEM and atomically resolved EELS. In that study[11] it was concluded that a Si-C$_3$ defect is non-planar due to a significantly better match between the measured and simulated EELS spectrum; however, the simulated spectrum of a flat Si-C$_4$ defect matched the experiment less well. Because of this discrepancy and the findings described above, we also simulated the EELS spectrum of the corrugated Si-C$_4$ graphene defect. FIG. 6 displays the simulated spectra of both the flat and the corrugated defect (normalized to the $\pi^*$ peak at 100 eV), overlaid on a new background-subtracted experimental signal that we have recorded with a higher dispersion than the original spectrum[11] (average of 20 spot spectra, with the Si-C$_4$ structure verified after acquisition by imaging); see Fig. S6 for the unprocessed spectrum).

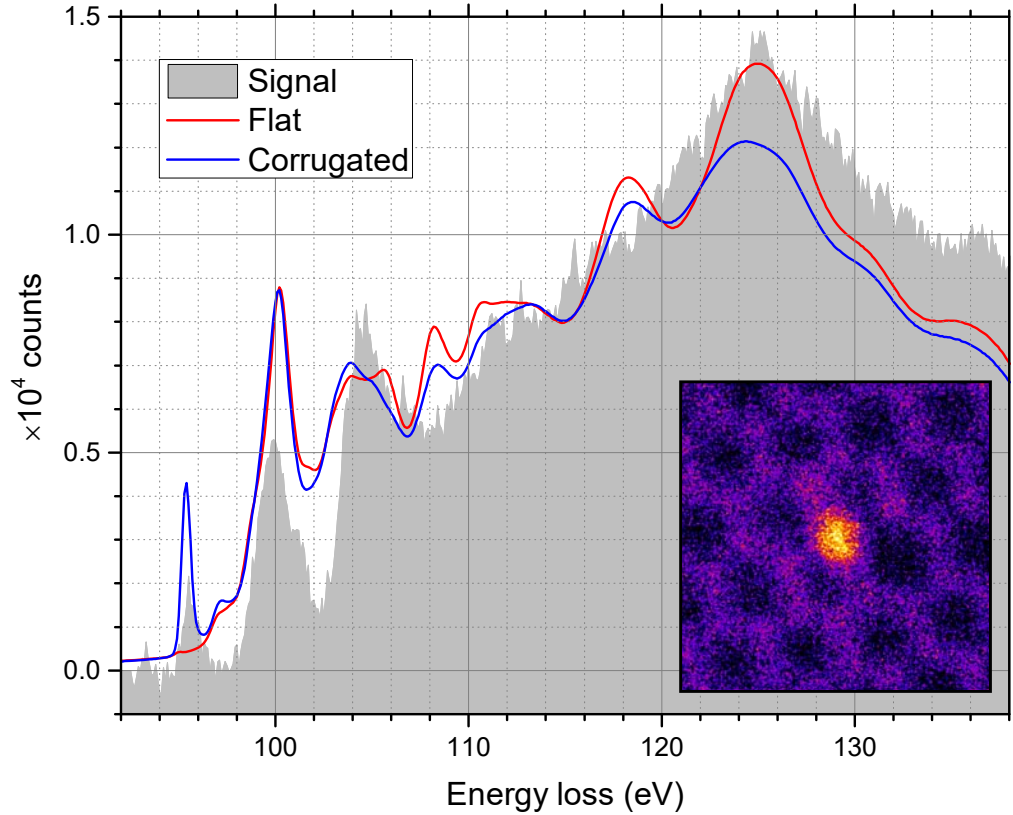

FIG. 6: Comparison of simulated and experimental EELS spectra of the Si-C$_4$ defect in graphene. Despite the out-of-plane distorted structure being the ground state, the originally proposed flat spectrum provides an overall better match to the experiment, apart from the small peak at ~96 eV that is not predicted for the flat structure. The inset shows a colored medium angle annular dark field STEM image of the Si-C$_4$ defect (the Si atom shows brighter due to its greater ability to scatter the imaging electrons).

The new spectrum closely resembles the originally reported one apart from a small additional peak around 96 eV. Comparing the recorded signal to the simulated spectra, the overall relative intensities of the $\pi^*$ and $\sigma^*$ contributions match the flat structure better, but both simulations overestimate the $\pi^*$ intensity at 100 eV. Intriguingly, despite the better overall match with the flat structure, the small pre-peak present in the new signal is only present in the spectrum simulated for the lowest energy corrugated structure, providing spectroscopic evidence for its existence. The lowest unoccupied states responsible for this peak are $\pi^*$ in character and form a „cross-like" pattern across the Si site, as shown in FIG. S7, markedly different from the lowest unoccupied states of the flat structure. However, considering the small energy barrier between the equivalent corrugated conformations discussed above, even at room temperature the spectrum may be integrated over a superposition of corrugated and nearly flat morphologies.

## IV. CONCLUSIONS

Density functional theory calculations (both restricted and unrestricted) were performed on pyrene-2C+Si and circumpyrene-2C+Si producing a variety of structures that were characterized using natural orbital occupations, natural bond orbital analyses, and molecular electrostatic potential plots. A non-planar low spin structure of $D_2$ symmetry was found to be the minimum for both cases, followed energetically by a low spin planar structure, 0.6 eV higher in energy for pyrene-2C+Si, but with a much smaller (only ~0.1 eV) energy difference for circumpyrene-2C+Si. A periodic graphene model with the non-planar defect had a 0.22 eV lower energy than a flat one. In the molecular calculations, the out-of-plane structure was shown to be a minimum by means of harmonic frequency calculations. The structures of the high spin state are much higher in energy than the minimum: about 0.85 eV higher for pyrene-2C+Si and about 1.3 eV higher for circumpyrene-2C+Si.

For pyrene-2C+Si, B3LYP and CAM-B3LYP were compared showing that there was not much difference either energetically, in the NBO characterization, or in the MEP plots. One feature of interest is that CAM-B3LYP shows greater open-shell character in the low spin case than was seen for B3LYP. This open-shell character demonstrates the variety of electronic structures which can be found in seemingly ordinary closed-shell cases.

NBO analyses showed the expected bonding configurations for the two structural symmetries present. The bonding of the silicon atom in the planar structure was found to have $sp^2d$ orbital character while in the non-planar structures, the bonding orbitals are $sp^3$ hybridized. The natural charges illustrated the charge transfer that occurs, demonstrating the buildup of positive charge at the silicon and the diffusion of negative charge into the pyrene-2C/cirumpyrene-2C structure, which was further evident in the MEP plots.

Despite carefully establishing the non-planar ground state nature of the Si-C$_4$ defect in graphene, electron energy loss spectra still seem to match the flat structure better, but imperfectly especially concerning a small peak at ~96 eV. Since the electrons used in transmission electron microscopy can impart large amounts of kinetic energy in addition to the available thermal energy, perhaps the defect is not able to stay in its ground state, but rather exists in a superposition of slightly different configurations that result in an average spectrum resembling the flat state.

## SUPPLEMENTARY MATERIAL

See Supplementary Material for NBO analysis, NPA charges, MEP for the pyrene-2C +Si structures, lowest unoccupied states of the graphene Si-C$_4$ defect, and Cartesian coordinates for the different optimized structures.

## ACKNOWLEDGMENTS

We are grateful to Michael Zehetbauer for continuous support and interest in this work and also thank Quentin Ramasse for many useful discussions. T.S. and J.K. acknowledge the Austrian Science Fund (FWF) for funding via projects P 28322-N36 and I 3181-N36. H.L. and T.S. are grateful to the Vienna Scientific Cluster (Austria) for computational resources (H.L.: project no. P70376). H.L. also acknowledges computer time at the computer cluster of the School for Pharmaceutical Science and Technology of the Tianjin University, China, T.P.H. acknowledges the Engineering and Physical Sciences Research Council (EPSRC) Doctoral Prize Fellowship that funded this research, and the ARC1 and ARC2 high-performance computing facilities at the University of Leeds.

Supplementary Information

# Structure and Electronic States of a Graphene Double Vacancy with an Embedded Si Dopant

Reed Nieman,[1] Adélia J. A. Aquino,[1,2] Trevor P. Hardcastle,[3,4] Jani Kotakoski,[5] Toma Susi,[5,*] and Hans Lischka[1,2,*]

[1] Department of Chemistry and Biochemistry, Texas Tech University Lubbock, TX 79409-1061, USA

[2] School of Pharmaceutical Sciences and Technology, Tianjin University, Tianjin, 300072 P.R.China

[3] SuperSTEM Laboratory, STFC Daresbury Campus, Daresbury WA4 4AD, United Kingdom

[4] University of Leeds, School of Chemical and Process Engineering, Faculty of Engineering, 211 Clarendon Rd, Leeds LS2 9JT, United Kingdom

[5] Faculty of Physics, University of Vienna, Boltzmanngasse 5, A-1090 Vienna, Austria

**Table of Contents**

TABLE SI: NBO bonding character analysis of pyrene-2C+Si using the CAM-B3LYP/6-311G(2d,1p) method. The carbon atom discussed is the carbon bonded to silicon.

| Structure | Bonding Character |
|---|---|
| | CAM-B3LYP/6-311G(2d,1p) |
| P4 ($D_2$) | $\alpha^a$: $0.53(sp^{2.98}d^{0.03})_{Si}+0.85(sp^{2.97}d^{0.01})_C$ |
| | $\beta^a$: $0.53(sp^{2.98}d^{1.03})_{Si}+0.85(sp^{2.98}d^{0.01})_C$ |
| P5 ($D_2$) | $\alpha^a$: $0.53(sp^{2.98}d^{0.03})_{Si}+0.85(sp^{3.00}d^{0.01})_C$ |
| | $\beta^a$: $0.53(sp^{2.98}d^{0.03})_{Si}+0.85(sp^{3.00}d^{0.01})_C$ |
| P6 ($D_2$) | $0.54(sp^{2.98}d^{0.02})_{Si}+0.84(sp^{3.20}d^{0.01})_C$ |
| P7 ($D_{2h}$) | $\alpha$: $0.45(sp^{2.01}d^{1.01})_{Si}+0.89(sp^{2.80}d^{0.01})_C$ |
| | $\beta$: $0.46(sp^{1.99}d^{0.99})_{Si}+0.89(sp^{2.78}d^{0.01})_C$ |
| P8 ($D_{2h}$) | $0.45(sp^{2.00}d^{1.00})_{Si}+0.89(sp^{2.92}d^{0.01})_C$ |
| P9 ($D_2$) | $\alpha$: $0.48(sp^{2.38}d^{0.62})_{Si}+0.88(sp^{2.39}d^{0.01})_C$ |
| | $\beta$: $0.48(sp^{2.92}d^{0.09})_{Si}+0.87(sp^{2.45}d^{0.01})_C$ |
| P10 ($D_{2h}$) | $\alpha$: $0.46(sp^{2.00}d^{1.00})_{Si}+0.89(sp^{2.41}d^{0.01})_C$ |
| | $\beta$: $0.45(sp^{2.00}d^{1.00})_{Si}+0.89(sp^{2.69}d^{0.01})_C$ |

[a] Averaged values of the four Si-C NBOs

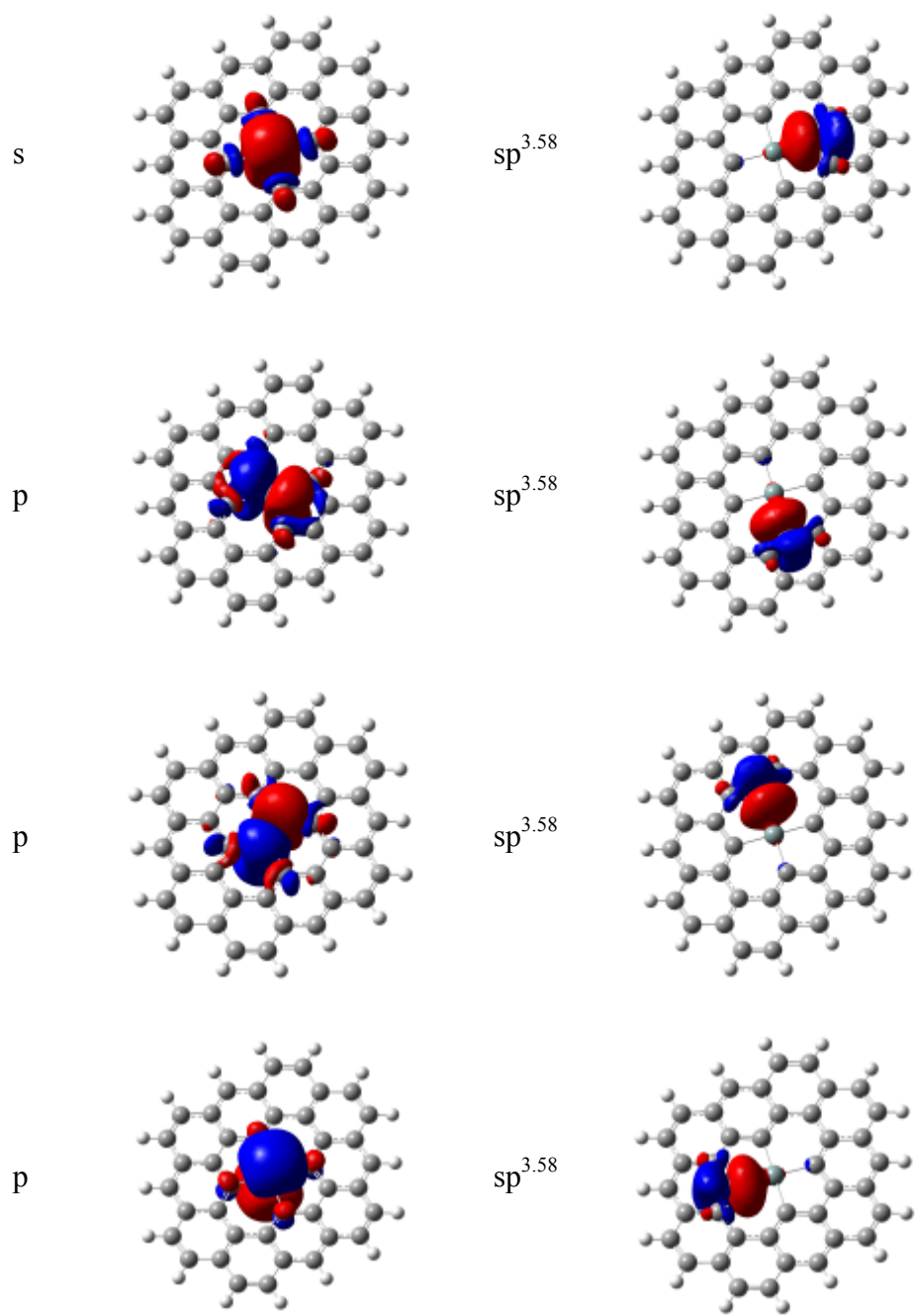

FIG. S1: NBO plots for circumpyrene-2C+Si using the B3LYP/6-311G(2d,1p) approach (CP2 $D_{2h}$). Isovalue = ±0.02 $e$/Bohr$^3$

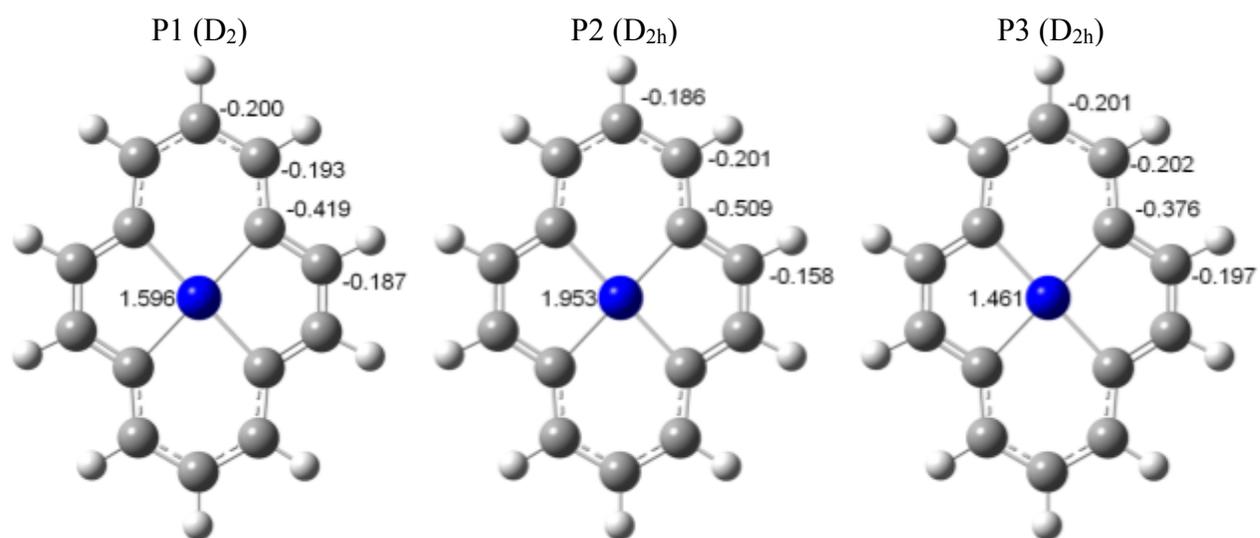

FIG. S2: Natural charges of carbon and silicon in units of $e$ of pyrene-2C+Si using the B3LYP/6-311G(2d,1p) method.

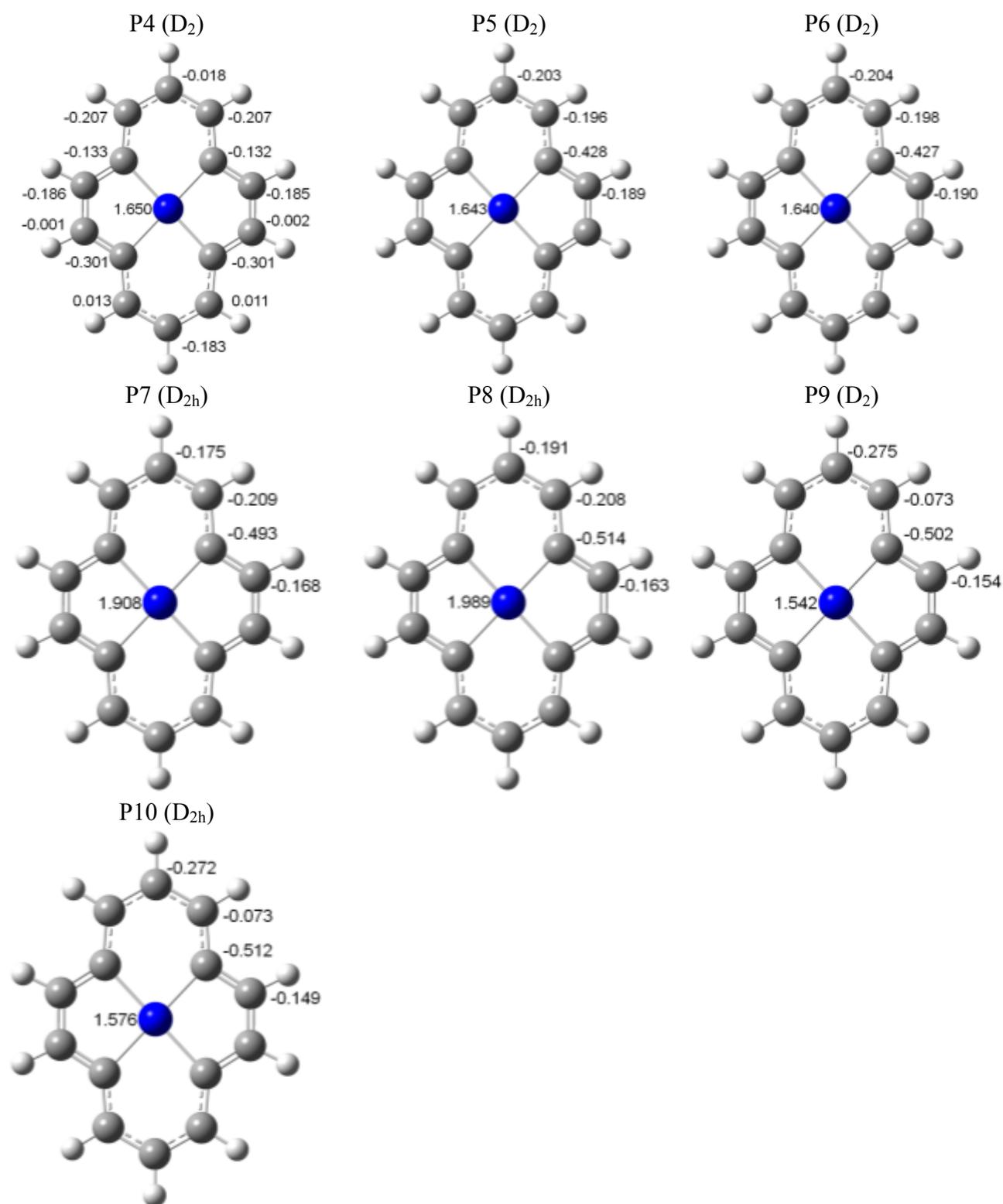

FIG. S3: Natural charges of carbon and silicon in units of *e* of pyrene-2C+Si using the CAM-B3LYP/6-311G(2d,1p) method.

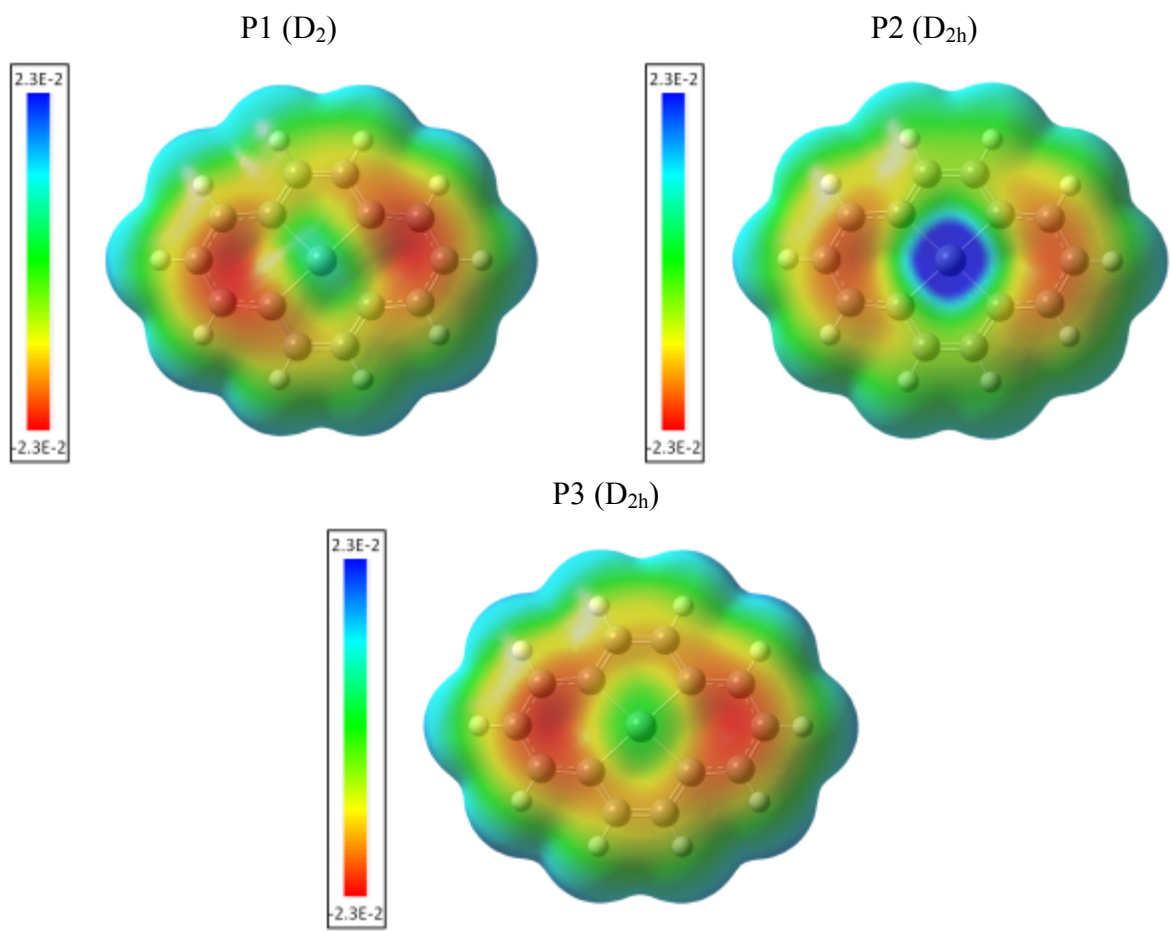

FIG. S4: Molecular electrostatic potential plot for pyrene-2C+Si mapped onto the electron density isosurface with an isovalue of 0.0004 $e$/Bohr$^3$ using the B3LYP/6-311G(2d,1p) method.

P4 (D$_2$)                      P5 (D$_2$)

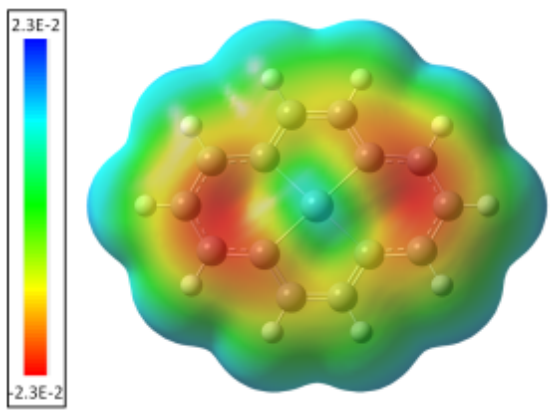 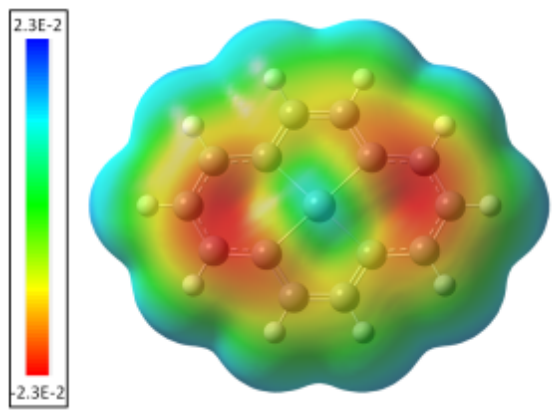

P6 (D$_2$)                      P7 (D$_{2h}$)

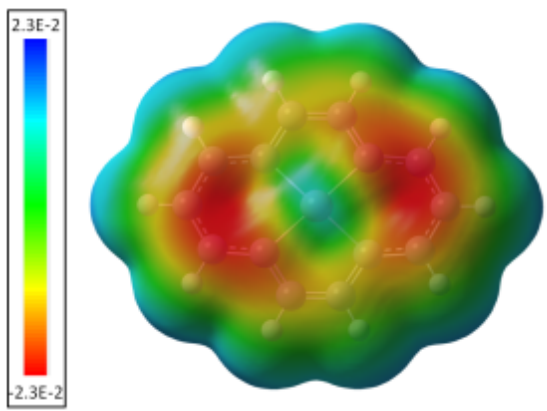 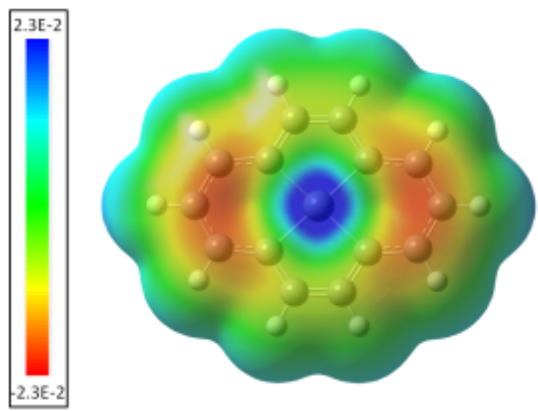

P8 (D$_{2h}$)                 P9 (D$_2$)

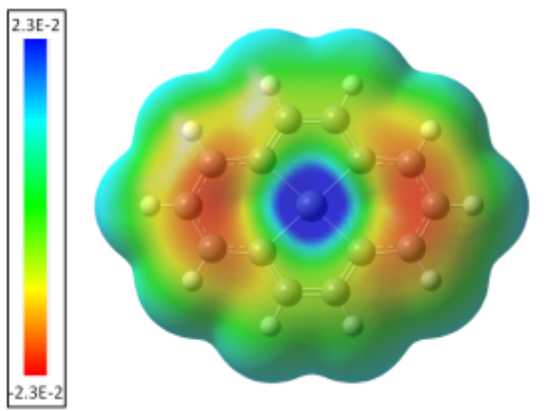 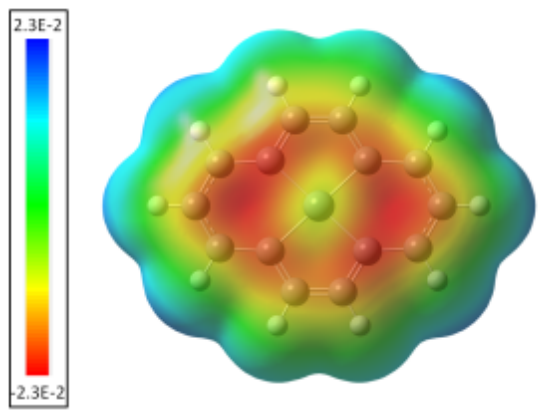

P10 (D$_{2h}$)

S8

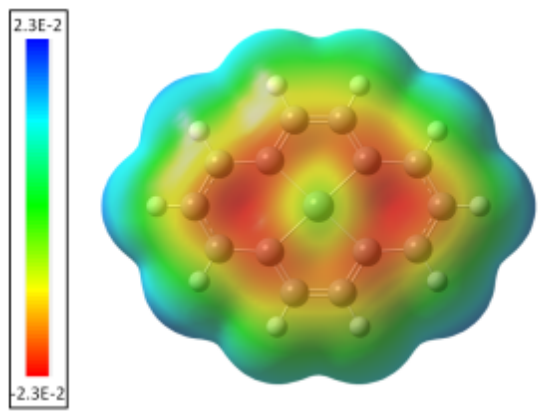

FIG. S5: Molecular electrostatic potential plot for pyrene-2C+Si mapped onto the electron density isosurface with an isovalue of 0.0004 $e$/Bohr$^3$ using the CAM-B3LYP/6-311G(2d,1p) method.

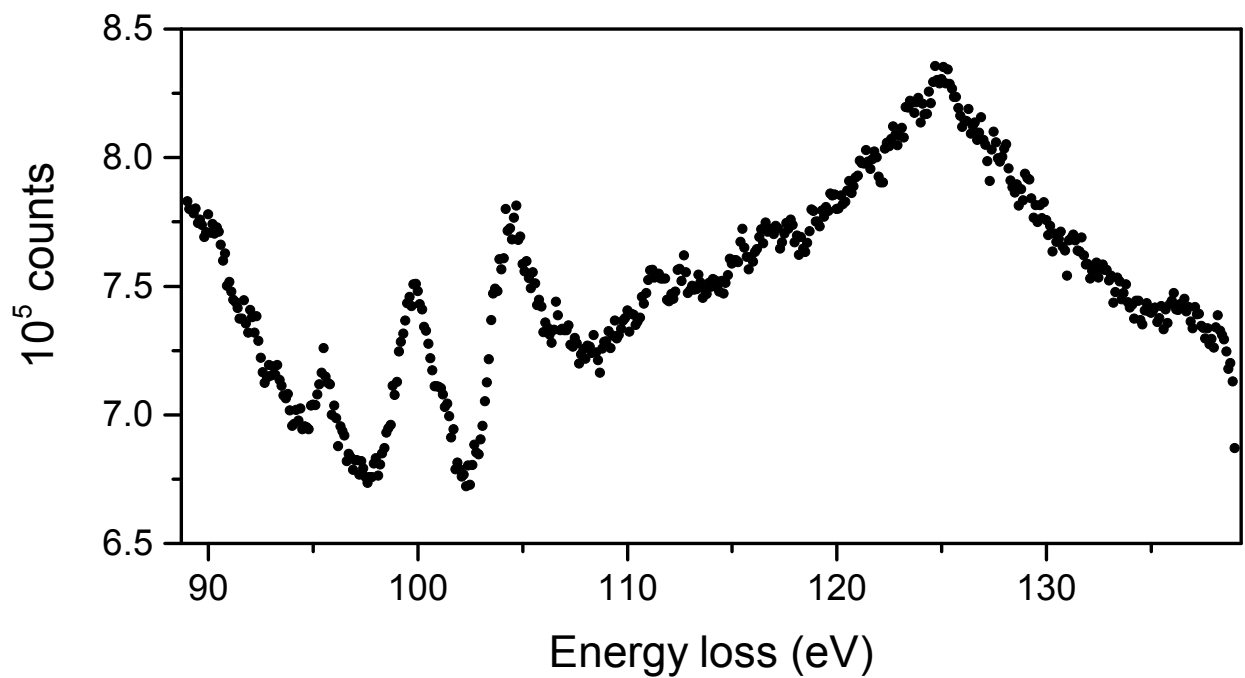

FIG. S6: Unprocessed EEL spectrum of the Si-C$_4$ site.

|      Flat      |   Corrugated   |
|----------------|----------------|

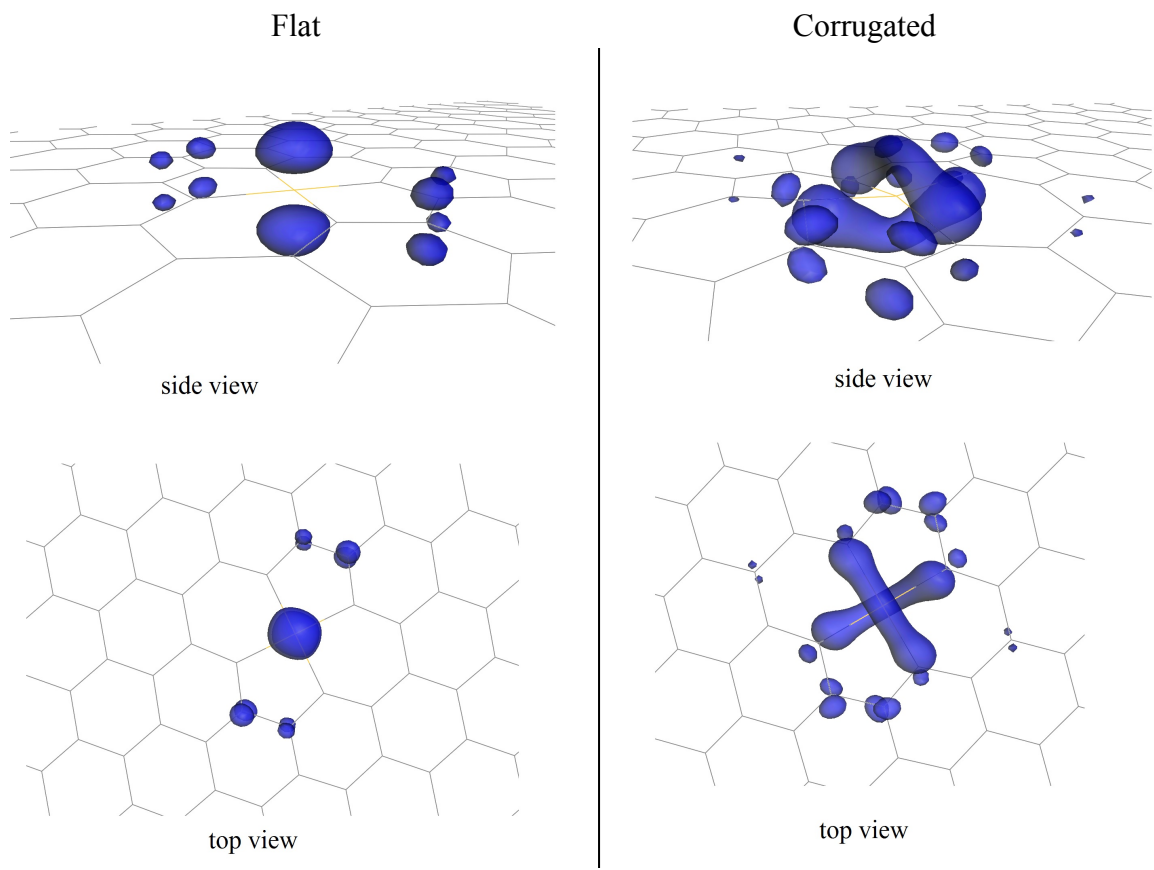

side view | side view
top view | top view

FIG. S7: First two unoccupied states immediately above the Fermi level spanning about 1 eV for the flat and corrugated Si/graphene structures from the periodic DFT calculations. Electron density isovalue is 0.277 $e$/Bohr$^3$.

Cartesian coordinates of structure P1 ($D_2$ symmetry) of pyrene-2C+Si using the B3LYP/6-311G(2d,1p) method

```
Si      0.0000213       0.0000977      -0.0000003
C       0.0000000       3.5245760      -0.0000071
C       0.0000000      -3.5245760       0.0000064
C       1.2114390       2.8334420      -0.0000058
C       1.2114390      -2.8334420       0.0000051
C      -1.2114390       2.8334420      -0.0000058
C      -1.2114390      -2.8334420       0.0000050
C       1.3108181       1.4163416      -0.1999923
C       1.3108198      -1.4163299       0.1999861
C      -1.3107757       1.4163473       0.1999391
C      -1.3107770      -1.4163356      -0.1999340
C       2.4650450       0.6811180      -0.0000017
C       2.4650450      -0.6811180       0.0000009
C      -2.4650450      -0.6811180       0.0000009
```

```
C   -2.4650450    0.6811180   -0.0000017
H    2.1451491   -3.3841955   -0.0807146
H   -3.4160765    1.1938400   -0.1659540
H   -3.4160312   -1.1938227    0.1660972
H    3.4160534    1.1938242    0.1660563
H   -0.0000023    4.6080724    0.0000150
H    2.1452243    3.3841978    0.0806401
H   -2.1452316    3.3841936   -0.0807018
H    3.4160082   -1.1938067   -0.1662018
H   -0.0000021   -4.6080168   -0.0000161
H   -2.1451566   -3.3841913    0.0807743
```

Cartesian coordinates of structure P2 ($D_{2h}$ symmetry) of pyrene-2C+Si using the B3LYP/6-311G(2d,1p) method

```
Si   -0.0000000    0.0000000    0.0000019
C     0.0000000    3.5245760    0.0000000
C     0.0000000   -3.5245760    0.0000000
C     1.2114390    2.8334420    0.0000000
C     1.2114390   -2.8334420    0.0000000
C    -1.2114390    2.8334420    0.0000000
C    -1.2114390   -2.8334420    0.0000000
C     1.3167860    1.4139350    0.0000117
C     1.3167860   -1.4139350   -0.0000108
C    -1.3167860    1.4139350    0.0000105
C    -1.3167860   -1.4139350   -0.0000094
C     2.4650450    0.6811180    0.0000000
C     2.4650450   -0.6811180    0.0000000
C    -2.4650450   -0.6811180    0.0000000
C    -2.4650450    0.6811180    0.0000000
H     2.1411325   -3.4009987   -0.0000000
H    -3.4449169    1.1713532    0.0000006
H    -3.4449169   -1.1713531   -0.0000011
H     3.4449169    1.1713531    0.0000003
H    -0.0000000    4.6076435   -0.0000024
H     2.1411325    3.4009987   -0.0000002
H    -2.1411325    3.4009987   -0.0000002
H     3.4449169   -1.1713531   -0.0000005
H    -0.0000000   -4.6076435    0.0000028
H    -2.1411325   -3.4009987   -0.0000002
```

Cartesian coordinates of structure P3 ($D_{2h}$ symmetry) of pyrene-2C+Si using the B3LYP/6-311G(2d,1p) method

```
Si    0.0000000    0.0003261   -0.0000009
C     0.0000000    3.5245602    0.0000000
C     0.0000000   -3.5245602    0.0000000
C     1.2114585    2.8335125    0.0000000
C     1.2114585   -2.8335125    0.0000000
C    -1.2114585    2.8335125    0.0000000
C    -1.2114585   -2.8335125    0.0000000
C     1.3048221    1.4261919    0.0000185
C     1.3047863   -1.4261979    0.0005887
C    -1.3048219    1.4261884   -0.0005890
C    -1.3047866   -1.4262014   -0.0000177
C     2.4650429    0.6810968    0.0000000
C     2.4650429   -0.6810968    0.0000000
C    -2.4650429   -0.6810968    0.0000000
C    -2.4650429    0.6810968    0.0000000
H     2.1478778   -3.3861020    0.0017420
H    -3.4290052    1.1957576    0.0013347
H    -3.4289618   -1.1958567   -0.0021052
H     3.4290050    1.1957557    0.0021015
H    -0.0000001    4.6071765   -0.0002465
H     2.1478378    3.3861214   -0.0026179
H    -2.1478392    3.3861223   -0.0017401
H     3.4289618   -1.1958587   -0.0013360
H     0.0000001   -4.6071979    0.0002465
H    -2.1478764   -3.3861013    0.0026195
```

Cartesian coordinates of structure P4 ($D_2$ symmetry) of pyrene-2C+Si using the B3LYP/6-311G(2d,1p) method

```
Si      0.000340     0.001167    -0.006564
C       0.000597     3.530312     0.003329
C       0.002019    -3.530758     0.004438
C       1.213381     2.828218     0.000054
C       1.211367    -2.828509    -0.002345
C      -1.213913     2.827539    -0.004376
C      -1.214342    -2.826714    -0.004169
C       1.308581     1.409550    -0.201836
C       1.307235    -1.408658     0.194738
C      -1.307328     1.410177     0.191136
C      -1.306678    -1.409890    -0.203215
C       2.463304     0.685869    -0.001279
C       2.464153    -0.685785     0.001276
C      -2.461781    -0.687243     0.001949
C      -2.464786     0.687070     0.001123
H       2.147036    -3.374294    -0.080698
H      -3.416327     1.198412    -0.151594
H      -3.409593    -1.201524     0.168301
H       3.412007     1.199269     0.163297
H       0.000763     4.612378     0.008883
H       2.148287     3.374076     0.085369
H      -2.148551     3.374386    -0.085969
H       3.413960    -1.198955    -0.156848
H       0.001507    -4.612798     0.007846
H      -2.149361    -3.372235     0.081261
```

Cartesian coordinates of structure P5 (*D*₂ symmetry) of pyrene-2C+Si using the B3LYP/6-311G(2d,1p) method

```
 Si      0.000004    0.000070    0.000010
 C      -0.000000    3.524558   -0.000000
 C      -0.000000   -3.524558    0.000000
 C       1.211457    2.833509   -0.000003
 C       1.211457   -2.833509    0.000003
 C      -1.211457    2.833509    0.000003
 C      -1.211457   -2.833509   -0.000003
 C       1.309496    1.412304   -0.195485
 C       1.309485   -1.412300    0.195546
 C      -1.309498    1.412300    0.195492
 C      -1.309485   -1.412294   -0.195545
 C       2.465043    0.681101    0.000000
 C       2.465043   -0.681101   -0.000000
 C      -2.465043   -0.681101    0.000000
 C      -2.465043    0.681101   -0.000000
 H       2.144627   -3.382532   -0.080984
 H      -3.414909    1.194207   -0.158781
 H      -3.414883   -1.194224    0.158856
 H       3.414912    1.194205    0.158772
 H      -0.000001    4.607088   -0.000003
 H       2.144615    3.382562    0.080957
 H      -2.144616    3.382560   -0.080957
 H       3.414884   -1.194223   -0.158859
 H      -0.000001   -4.607090   -0.000001
 H      -2.144629   -3.382530    0.080975
```

Cartesian coordinates of structure P6 ($D_2$ symmetry) of pyrene-2C+Si using the B3LYP/6-311G(2d,1p) method

```
Si         0.000003     0.000076     0.000009
C          0.000000     3.524560     0.000000
C          0.000000    -3.524560     0.000000
C          1.211458     2.833513     0.000000
C          1.211458    -2.833513     0.000000
C         -1.211458     2.833513     0.000000
C         -1.211458    -2.833513     0.000000
C          1.309250     1.414896    -0.195809
C          1.309238    -1.414892     0.195873
C         -1.309252     1.414892     0.195816
C         -1.309238    -1.414887    -0.195873
C          2.465043     0.681097     0.000000
C          2.465043    -0.681097     0.000000
C         -2.465043    -0.681097     0.000000
C         -2.465043     0.681097     0.000000
H          2.144797    -3.382413    -0.079930
H         -3.415083     1.193713    -0.159203
H         -3.415053    -1.193735     0.159282
H          3.415085     1.193712     0.159195
H         -0.000001     4.607094    -0.000002
H          2.144782     3.382447     0.079902
H         -2.144783     3.382445    -0.079901
H          3.415054    -1.193734    -0.159284
H         -0.000001    -4.607095    -0.000001
H         -2.144798    -3.382411     0.079921
```

Cartesian coordinates of structure P7 ($D_{2h}$ symmetry) of pyrene-2C+Si using the B3LYP/6-311G(2d,1p) method

```
Si        -0.000000     0.000000    -0.000000
C          0.000000     3.524576     0.000000
C          0.000000    -3.524576     0.000000
C          1.211439     2.833442     0.000000
C          1.211439    -2.833442     0.000000
C         -1.211439     2.833442     0.000000
C         -1.211439    -2.833442     0.000000
C          1.308024     1.409936    -0.000000
C          1.308024    -1.409936    -0.000000
C         -1.308024     1.409936    -0.000000
C         -1.308024    -1.409936    -0.000000
C          2.465045     0.681118     0.000000
C          2.465045    -0.681118     0.000000
C         -2.465045    -0.681118     0.000000
C         -2.465045     0.681118     0.000000
H          2.141400    -3.397000     0.000000
H         -3.440737     1.174141     0.000000
H         -3.440737    -1.174141     0.000000
H          3.440737     1.174141     0.000000
H         -0.000000     4.606865     0.000000
H          2.141400     3.397000     0.000000
H         -2.141400     3.397000     0.000000
H          3.440737    -1.174141     0.000000
H          0.000000    -4.606865    -0.000000
H         -2.141400    -3.397000     0.000000
```

Cartesian coordinates of structure P8 ($D_{2h}$ symmetry) of pyrene-2C+Si using the B3LYP/6-311G(2d,1p) method

```
 Si        0.000000   -0.000000   -0.000000
 C         0.000000    3.524576    0.000000
 C         0.000000   -3.524576    0.000000
 C         1.211439    2.833442    0.000000
 C         1.211439   -2.833442    0.000000
 C        -1.211439    2.833442    0.000000
 C        -1.211439   -2.833442    0.000000
 C         1.315650    1.412869   -0.000000
 C         1.315650   -1.412869   -0.000000
 C        -1.315650    1.412869   -0.000000
 C        -1.315650   -1.412869   -0.000000
 C         2.465045    0.681118    0.000000
 C         2.465045   -0.681118    0.000000
 C        -2.465045   -0.681118    0.000000
 C        -2.465045    0.681118    0.000000
 H         2.140509   -3.399025   -0.000000
 H        -3.442135    1.172375   -0.000000
 H        -3.442135   -1.172375   -0.000000
 H         3.442135    1.172375   -0.000000
 H         0.000000    4.606772   -0.000000
 H         2.140509    3.399025   -0.000000
 H        -2.140509    3.399025   -0.000000
 H         3.442135   -1.172375    0.000000
 H        -0.000000   -4.606772   -0.000000
 H        -2.140509   -3.399025    0.000000
```

Cartesian coordinates of structure P9 (*D*₂ symmetry) of pyrene-2C+Si using the B3LYP/6-311G(2d,1p) method

```
Si        -0.000000     0.000001     0.000031
C         -3.524761     0.000090    -0.000045
C          3.524760    -0.000090     0.000048
C         -2.833417     1.211325     0.000038
C          2.833495     1.211204    -0.000015
C         -2.833496    -1.211205     0.000018
C          2.833415    -1.211323    -0.000035
C         -1.386698     1.297539     0.071993
C          1.386822     1.297542    -0.071952
C         -1.386822    -1.297545    -0.071914
C          1.386694    -1.297538     0.071932
C         -0.680991     2.465075     0.000008
C          0.681142     2.465135    -0.000013
C          0.680989    -2.465073    -0.000003
C         -0.681143    -2.465137     0.000019
H          3.393546     2.143163     0.025713
H         -1.184238    -3.433713     0.063476
H          1.184122    -3.433634    -0.063499
H         -1.184123     3.433634    -0.063512
H         -4.606858     0.000130    -0.000074
H         -3.393381     2.143334    -0.025648
H         -3.393545    -2.143163     0.025811
H          1.184238     3.433711     0.063436
H          4.606857    -0.000128     0.000045
H          3.393375    -2.143332    -0.025814
```

Cartesian coordinates of structure P10 ($D_{2h}$ symmetry) of pyrene-2C+Si using the B3LYP/6-311G(2d,1p) method

```
Si         0.000000    -0.000008    -0.000012
C          0.000004     3.524760    -0.000047
C         -0.000004    -3.524761     0.000046
C          1.211256     2.833446     0.000036
C          1.211273    -2.833466    -0.000017
C         -1.211274     2.833466     0.000016
C         -1.211254    -2.833445    -0.000037
C          1.299085     1.386187     0.000402
C          1.299154    -1.386274    -0.000376
C         -1.299158     1.386276    -0.000048
C         -1.299081    -1.386184    -0.000003
C          2.465058     0.681051     0.000006
C          2.465152    -0.681082    -0.000015
C         -2.465056    -0.681050    -0.000005
C         -2.465154     0.681082     0.000017
H          2.142335    -3.396294     0.000082
H         -3.438009     1.180869    -0.000149
H         -3.437871    -1.180933     0.000135
H          3.437879     1.180927    -0.000452
H          0.000021     4.606843    -0.000228
H          2.142345     3.396226    -0.000090
H         -2.142335     3.396298     0.000038
H          3.438002    -1.180875     0.000529
H         -0.000021    -4.606844     0.000243
H         -2.142345    -3.396221    -0.000057
```

Cartesian coordinates of structure CP1 ($D_2$ symmetry) of circumpyrene-2C+Si using the B3LYP/6-311G(2d,1p) method

```
Si     0.0000082   -0.0000000    0.0000000
C      3.4996422    0.0000002    0.0000013
C     -3.4996567    0.0000003   -0.0000012
C      2.7787776    1.2513988   -0.0926702
C     -2.7787795    1.2513988    0.0926851
C      2.7787776   -1.2513985    0.0926718
C     -2.7787795   -1.2513985   -0.0926867
C      1.3551457    1.3297097   -0.2044730
C     -1.3551405    1.3297067    0.2044940
C      1.3551456   -1.3297097    0.2044741
C     -1.3551404   -1.3297066   -0.2044951
C      0.7268927    2.5532891   -0.0566937
C     -0.7268914    2.5532860    0.0567304
C     -0.7268914   -2.5532859   -0.0567306
C      0.7268927   -2.5532891    0.0566940
C      5.6480814    1.2560828    0.0000000
C      5.6480814   -1.2560828    0.0000000
C     -5.6480814   -1.2560828    0.0000000
C     -5.6480814    1.2560828    0.0000000
C      3.5414432    2.4817206   -0.0287623
C      3.5414432   -2.4817205    0.0287629
C     -3.5414490   -2.4817281   -0.0287694
C     -3.5414490    2.4817282    0.0287688
C     -4.9925337    2.4482104    0.0000000
C      4.9925337    2.4482104    0.0000000
C      4.9925337   -2.4482104    0.0000000
C     -4.9925337   -2.4482104    0.0000000
C      1.4547230   -3.7535512    0.0040261
C     -1.4547221   -3.7535549   -0.0040435
C     -1.4547221    3.7535549    0.0040434
C      1.4547230    3.7535512   -0.0040259
C     -2.8452352   -3.7019986    0.0000000
C     -2.8452352    3.7019986    0.0000000
C      2.8452352   -3.7019986    0.0000000
C      2.8452352    3.7019986    0.0000000
C      0.6850947    4.9452338    0.0000000
C     -0.6850947    4.9452338    0.0000000
C     -0.6850947   -4.9452338    0.0000000
C      0.6850947   -4.9452338    0.0000000
C      4.9238436   -0.0000001   -0.0000002
C     -4.9238488   -0.0000001    0.0000002
H      5.5314381    3.3873835    0.0337344
```

```
H         5.5314381     -3.3873836     -0.0337343
H        -5.5314368      3.3873818     -0.0337398
H        -5.5314368     -3.3873819      0.0337397
H         6.7314129      1.2201897      0.0094780
H         6.7314130     -1.2201897     -0.0094784
H        -6.7314111     -1.2201896      0.0094816
H        -6.7314111      1.2201896     -0.0094812
H         3.4101284     -4.6288133     -0.0159644
H        -3.4101281     -4.6288107      0.0159892
H        -3.4101280      4.6288108     -0.0159888
H         3.4101284      4.6288133      0.0159639
H         1.2037538     -5.8987949     -0.0113360
H        -1.2037539     -5.8987942      0.0113634
H         1.2037538      5.8987949      0.0113359
H        -1.2037538      5.8987942     -0.0113634
```

Cartesian coordinates of structure CP2 ($D_{2h}$ symmetry) of circumpyrene-2C+Si using the B3LYP/6-311G(2d,1p) method

```
Si     0.0000000     0.0000001     0.0000019
C      3.4989334     0.0000005     0.0000024
C     -3.4989334     0.0000007     0.0000023
C      2.7788566     1.2551223    -0.0000660
C     -2.7788565     1.2551222     0.0000691
C      2.7788566    -1.2551214     0.0000693
C     -2.7788565    -1.2551218    -0.0000662
C      1.3562603     1.3339174    -0.0002123
C     -1.3562604     1.3339177     0.0002152
C      1.3562603    -1.3339179     0.0002151
C     -1.3562606    -1.3339167    -0.0002125
C      0.7311916     2.5623098    -0.0000428
C     -0.7311915     2.5623090     0.0000428
C     -0.7311914    -2.5623104    -0.0000430
C      0.7311914    -2.5623082     0.0000428
C      5.6480814     1.2560828     0.0000000
C      5.6480814    -1.2560828     0.0000000
C     -5.6480814    -1.2560828     0.0000000
C     -5.6480814     1.2560828     0.0000000
C      3.5419197     2.4842562    -0.0000027
C      3.5419197    -2.4842560     0.0000038
C     -3.5419197    -2.4842558    -0.0000028
C     -3.5419198     2.4842562     0.0000036
C     -4.9925337     2.4482104     0.0000000
C      4.9925337     2.4482104     0.0000000
C      4.9925337    -2.4482104     0.0000000
C     -4.9925337    -2.4482104     0.0000000
C      1.4523498    -3.7580913    -0.0000170
C     -1.4523495    -3.7580903     0.0000169
C     -1.4523497     3.7580911    -0.0000170
C      1.4523496     3.7580907     0.0000170
C     -2.8452352    -3.7019986     0.0000000
C     -2.8452352     3.7019986     0.0000000
C      2.8452352    -3.7019986     0.0000000
C      2.8452352     3.7019986     0.0000000
C      0.6850947     4.9452338     0.0000000
C     -0.6850947     4.9452338     0.0000000
C     -0.6850947    -4.9452338     0.0000000
C      0.6850947    -4.9452338     0.0000000
C      4.9214655    -0.0000003    -0.0000090
C     -4.9214656    -0.0000004    -0.0000090
H      5.5355671     3.3859285     0.0000214
```

```
H      5.5355669   -3.3859286   -0.0000170
H     -5.5355670    3.3859285   -0.0000171
H     -5.5355669   -3.3859286    0.0000214
H      6.7312853    1.2197858   -0.0000087
H      6.7312853   -1.2197857    0.0000049
H     -6.7312853   -1.2197857   -0.0000087
H     -6.7312853    1.2197858    0.0000053
H      3.4112864   -4.6287297   -0.0000007
H     -3.4112865   -4.6287297    0.0000005
H     -3.4112863    4.6287298   -0.0000007
H      3.4112863    4.6287298    0.0000004
H      1.2003938   -5.9010537   -0.0000112
H     -1.2003938   -5.9010537    0.0000111
H      1.2003937    5.9010537    0.0000111
H     -1.2003938    5.9010537   -0.0000112
```

Cartesian coordinates of structure CP3 ($D_{2h}$ symmetry) of circumpyrene-2C+Si using the B3LYP/6-311G(2d,1p) method

```
Si     0.0000442     0.0000000    -0.0000003
C      3.5102074     0.0000005    -0.0000586
C     -3.5102042    -0.0000005     0.0000586
C      2.7889221     1.2459761     0.0000323
C     -2.7889154     1.2459753     0.0001310
C      2.7889222    -1.2459754    -0.0001308
C     -2.7889153    -1.2459758    -0.0000325
C      1.3725877     1.3312538     0.0001390
C     -1.3725931     1.3312238     0.0002136
C      1.3725877    -1.3312536    -0.0002131
C     -1.3725931    -1.3312240    -0.0001397
C      0.7152835     2.5589160     0.0001873
C     -0.7152747     2.5589464     0.0002161
C     -0.7152745    -2.5589464    -0.0001872
C      0.7152836    -2.5589160    -0.0002159
C      5.6480814     1.2560828     0.0000000
C      5.6480814    -1.2560828     0.0000000
C     -5.6480814    -1.2560828     0.0000000
C     -5.6480814     1.2560828     0.0000000
C      3.5450598     2.4835855     0.0000214
C      3.5450598    -2.4835853    -0.0000589
C     -3.5450583    -2.4835828    -0.0000215
C     -3.5450582     2.4835825     0.0000590
C     -4.9925337     2.4482104     0.0000000
C      4.9925337     2.4482104     0.0000000
C      4.9925337    -2.4482104     0.0000000
C     -4.9925337    -2.4482104     0.0000000
C      1.4541259    -3.7535817    -0.0000994
C     -1.4541154    -3.7535545    -0.0000843
C     -1.4541155     3.7535545     0.0000993
C      1.4541258     3.7535817     0.0000845
C     -2.8452352    -3.7019986     0.0000000
C     -2.8452352     3.7019986     0.0000000
C      2.8452352    -3.7019986     0.0000000
C      2.8452352     3.7019986     0.0000000
C      0.6850947     4.9452338     0.0000000
C     -0.6850947     4.9452338     0.0000000
C     -0.6850947    -4.9452338     0.0000000
C      0.6850947    -4.9452338     0.0000000
C      4.9340030    -0.0000003    -0.0000281
C     -4.9340034     0.0000003     0.0000281
H      5.5333236     3.3867429     0.0000010
H      5.5333235    -3.3867429     0.0000191
```

```
H     -5.5333216     3.3867429    -0.0000190
H     -5.5333217    -3.3867429    -0.0000011
H      6.7317573     1.2245980     0.0000076
H      6.7317572    -1.2245978     0.0000157
H     -6.7317570    -1.2246003    -0.0000076
H     -6.7317570     1.2246001    -0.0000157
H      3.4087110    -4.6292851     0.0000907
H     -3.4087062    -4.6292864     0.0000550
H     -3.4087063     4.6292863    -0.0000907
H      3.4087108     4.6292852    -0.0000550
H      1.2050100    -5.8980511     0.0000532
H     -1.2050169    -5.8980462     0.0000489
H      1.2050100     5.8980511    -0.0000490
H     -1.2050168     5.8980462    -0.0000531
```